\documentclass[doublecol]{epl2} 

\usepackage{graphicx}
\usepackage{amsmath}
\usepackage{url}
\usepackage{longtable}

\newcommand{\zmax}{z_{\textrm{max}}}
\newcommand{\zmin}{z_{\textrm{min}}}

\newcommand{\beq}{\begin{equation}}
\newcommand{\eeq}{\end{equation}}
\newcommand{\beqNo}{\begin{equation*}}
\newcommand{\eeqNo}{\end{equation*}}

\newcommand{\kf}{k_f}
\newcommand{\df}{d_f}

\title{Morphology-dependent random binary fragmentation of in silico fractal-like agglomerates}
\shorttitle{Random binary fragmentation of \textit{in silico} fractal-like agglomerates} 

\author{Y. Drossinos\inst{1}\thanks{ E-mail: \email{yannis.drossinos@ec.europa.eu} (corresponding author)}
\and A. D. Melas\inst{2} \and M. Kostoglou\inst{3}
\and L. Isella\inst{4}}
\shortauthor{Y. Drossinos \etal}

\institute{                    
  \inst{1} European Commission, Joint Research Centre, 21027 Ispra (VA), Italy \\
  \inst{2} Aerosol \& Particle Technology Laboratory, CPERI/CERTH, 57001
   Thessaloniki, Greece \\
  \inst{3} Department, of Chemistry, Aristotle University, 54124
   Thessaloniki, Greece \\
  \inst{4} European Commission, DG Trade, 1040 Bruxelles, Belgium \\
}
 {Fracture/brittleness}

\abstract{
Linear binary fragmentation of synthetic fractal-like agglomerates composed of spherical, equal-size, touching monomers is numerically investigated. Agglomerates of different morphologies are fragmented via random bond removal. The fragmentation algorithm relies on mapping each agglomerate onto an adjacency matrix. The numerically-determined fragment size distributions are U-shaped, clusters break predominantly into two largely dissimilar fragments, becoming more uniform as the fractal dimension decreases. A
symmetric beta distribution reproduces the fragment distribution rather accurately.
Its exponent depends on the structure (fractal dimension) and number of monomers of the initial agglomerate. A universal fragment distribution, a function only
of the initial fractal dimension, is derived by requiring that it satisfy the fragmentation conversation laws and the straight-chain limit. We argue that the fragmentation rate is proportional to the initial agglomerate size.
}

\begin{document}

\maketitle

\section{Introduction}

Fragmentation is important in nature and industry. Polymer degradation~\cite{Montroll1940},
fracture of solids and volcanic eruptions~\cite{Turcotte}, network topology and resilience~\cite{FragRandom2015}, 
droplet~\cite{Lasheras} and agglomerate~\cite{Mike} breakup
in turbulent flow are just a few examples where fragmentation
determines the outcome of a process.

The fragmentation of linear chains has been
extensively studied in polymer science
since bond degradation or depolymerization
breaks up the polymer.
In most cases random scission~\cite{Montroll1940,ZiffMcGrady1985,Krapivsky1994},
whereby all bonds break with equal probability, 
is considered. In colloids,
the size distribution of colloidal
aggregates in a fluid flow, and thence the
suspension rheology, is greatly influenced
by hydrodynamic shear-induced stresses that may break or restructure
the particles~\cite{VanniGastaldi2011}.

In aerosol science, which plays
a fundamental role in \textit{e.g.,} 
atmospheric sciences (cloud formation),
air pollution (ultimate fate of atmospheric particulate pollutants),
and industrial production of pharmaceuticals, the particle size 
distribution is one of the most important parameters that
specifies the chemical and physical
properties of the aerosol.

The dynamics of the size distribution,
as determined by agglomeration, whereby the colliding particles
retain their identity (or coagulation whereby colliding particle
coalesce) and fragmentation, is usually described by  mean-field rate equations.
These population balance equations are
generalizations of the Smoluchowski coagulation equation~\cite{Smoluchowski}.
The kinetics of irreversible agglomeration
has been extensively analysed, often with emphasis on scaling properties
of the size distribution or the characteristic cluster size~\cite{Meakin1988}.
As clusters grow their number decreases continuously, their characteristic size increases,
and the size distribution shrinks.
In contrast, if fragmentation occurs, a likely event as the ever-growing clusters are more
likely to fragment,
a steady-state distribution may be established
in the long-time limit.

Our primary interest in this work lies in the characterization of the
size distribution of fragments (and the associated
fragmentation kernel)
upon random scission of fractal-like agglomerates,
and its dependence on the morphology
of the initial agglomerate, as described by its fractal dimension (for fixed
fractal prefactor).
We consider numerically the linear, binary fragmentation of synthetic,
loop-less, fractal-like agglomerates upon random link removal. Agglomerates
that satisfy a scaling law over a finite range (fractal-like) are generated
\emph{in silico}.
They are composed of equal-size, point-touching spherical monomers.

\section{Discrete fragmentation kernel}
\label{sec:FragKernel}

Consider the irreversible fragmentation equation~\cite{McGradyZiff1987},
for the discrete particle size distribution
$n_i (t)$, the number of agglomerates composed of $i$ monomers (of size $i$)
per unit volume at time $t$
\beq
\frac{d n_i (t)}{dt} = \sum_{j = i + 1}^{i_{\textrm{max}}} \, a_j \, p_{ij} \, n_j(t) - a_i \, n_i(t) ,
\label{eq:FragGDE}
\eeq
where $a_i$ is the (herein taken to be time-independent) fragmentation rate of a particle
of size $i$, and $p_{ij}$ is the distribution
of fragments of size $i$ resulting from the break-up of a particle of 
initial size $j$. The maximum particle size $i_{\textrm{max}}$ is usually
taken to be infinity.


The fragment size distribution must obey a number of 
conservation laws~\cite{McGradyZiff1987, Kostoglou1997}. The first
constraint gives the expected number of fragments upon 
single-bond removal.
For binary fragmentation it reduces to
\beq
\sum_{i=1}^{j-1} p_{ij} = 2, \quad \quad \forall j .
\label{eq:ConservationLaw1Discrete}
\eeq
Since each particle breaks into two fragments, the fragment
size distribution must be symmetric, $p_{ij} = p_{(j-i)j}$.
The second conservation law is mass conservation:
the sum of the number of monomers in all the fragments must equal
the number of monomers in the initial agglomerate,
\beq
\sum_{i=1}^{j-1} i \, p_{ij} = j, \quad \quad \forall j .
\label{eq:ConservationLaw2Discrete}
\eeq
In Appendix A (Supplementary Material) we 
show that for binary fragmentation eq.~(\ref{eq:ConservationLaw1Discrete})
implies eq.~(\ref{eq:ConservationLaw2Discrete}),
rendering only one conservation
law independent.
In addition, 
eq.~(\ref{eq:ConservationLaw1Discrete}) suggests that $g_{ij} = p_{ij}/2$ defines
a (discrete) probability distribution. Then, 
eq.~(\ref{eq:ConservationLaw2Discrete}) shows that the average
fragment size is $j/2$ (as expected).

The fragmentation rate $a_j$ is estimated by assuming that all
bonds break with equal probability. Let the lifetime of each bond be
characterized by a probability distribution, \emph{e.g.,} a Poisson distribution
with average bond lifetime (characteristic time of fragmentation) $\tau$.
The number of bonds in an $j$-mer that would break in time $t = \tau$ is
the number of monomers times the probability of a bond breaking. Thus, the fragmentation
frequency is proportional to the initial size of the agglomerate, $a_j = j/\tau$.

\section{Continuous fragmentation kernel}

It is occasionally convenient to approximate the discrete particle size distribution
by a continuous distribution. This approximation is very accurate for $j \gg 1$.
We also introduce a vector of morphological descriptors $\vect{s}$ to account
for the morphology of the aggregate~\cite{Margaritis2001}. Accordingly, the discrete
size distribution $n_i(t)$ is replaced by the continuous distribution
$n(x,t; \vect{s})$ where $n(x,t; \vect{s}) \upd x$ is the number of particles per unit volume
in the size range $x$ and $x + \upd x$ and at morphological state $\vect{s}$.
For the cases considered herein, the additional morphological descriptor is
the fractal dimension $\vect{s} = d_f$. 
The discrete variables
$i$ and $j$ are replaced by the continuous variables $x$ (fragment size)
and $y$ (initial agglomerate size), respectively. The continuous
version of eq.~(\ref{eq:FragGDE}) becomes
\begin{align}
\frac{d n (x,t;d_f)}{dt} = \int_x^{x_{\textrm{max}}} \,
& \upd y \,  a(y) \, p (x,y; d_f) \, n(y,t; d_f) \nonumber \\
& - a(x) \, n(x,t; d_f) ,
\label{eq:FragGDECont}
\end{align}
where $p(x,y; d_f)$ is the fragmentation density function to obtain a $x$-mer from the
fragmentation of a $y$-mer. It is related to its discrete counterpart via
$p(x,y; d_f) \upd x = p_{ij}$.
The fragmentation frequency is $a(x) = x/\tau$.
The fragmentation density function must satisfy
the conservation laws (see, also, Appendix A, Supplementary Material)
\beq
\int_0^y \, \upd x \, p(x,y; d_f) = 2;   \quad 
 \int_0^y \upd x \, x \, p(x,y; d_f) = y .
\label{eq:ConsrvationLawsCont}
\eeq
with the symmetry condition $p(x,y; d_f) = p(y-x,y; d_f)$. As in the discrete
case, only one  conservation law is independent. The function $p(x,y;d_f)$ is defined
for $1 \le x \le y-1$. 
%
Henceforth, we will present results for the continuous fragment size distribution.
The discrete case is summarized in Appendix C (Supplementary Material). 
%

\section{Generation and fragmentation of synthetic agglomerates}
\label{sec:FragAgglomerates}

We generated \textit{in silico} independent, fractal-like agglomerates of
specified structure (fixed fractal prefactor $k_f = 1.3$ and
varying dimension $d_f = 1.6, 1.8, 2.1$) and mass (number of monomers $j$).
We used the hierarchical
tunable, off-lattice, cluster-cluster agglomeration algorithm,
initially proposed by Thouy and Jullien~\cite{ThouyJullien} and later
modified by
Filippov et al.~\cite{FilippovAlgo}, to create
mostly loop-less, agglomerates composed
of equal-sized, spherical, point-touching monomers.
We only considered loop-less agglomerates.
The algorithm  does not aim to reproduce a
physical agglomeration mechanism in that the motion of
the colliding clusters is not explicitly simulated.
Instead, it is based on geometric (including steric)
considerations. The advantage
of such a geometric algorithm is that the desired cluster
morphology ($\df, \kf$) is an input (whereas in explicit
cluster simulations it is an output).
The synthetic agglomerates satisfy by construction
exactly the fractal-like scaling law
$j = k_f \, (R_g/R_1)^{d_f}$
where $R_g$ is the radius of gyration of the cluster~\cite{Melas2014}
and $R_1$ is the monomer radius.
As summarized in Appendix B (table~\ref{table:Agglomerates}, Supplementary Material)
the number of agglomerates varied from 400,000 to 25,000.
The number of monomers
per cluster varied from approximately 80 to 800.

According to the cluster generation algorithm
(Appendix B, Supplementary Material,)
clusters are created
via monomer-monomer contact.
If the resulting cluster
does not have loops, 
the addition of a cluster onto an existing cluster generates only one bond. 
Thus, the total number of bonds in a loop-less $j$-cluster is $j-1$;
the  sum of bonds connecting
all monomers in a cluster [over both pairs  ($i,j-i$) and ($j-i,i$)]
is $2(j-1)$.
If $f_{ij}$ is the number of bonds that upon breaking would
divide the initial cluster into two fragments of sizes $i$ and $j-i$,
then their sum is $\sum_{i=1}^{j-1} f_{ij} = 2 (j-1)$,
as discussed extensively in Ref.\cite{Odriozola2002}.
The corresponding fragment distribution becomes $p_{ij} = f_{ij}/(j-1)$.
Equivalently, these clusters
have a (mean) coordination number
of $c_j = 2(j-1)/j$\cite{GastaldiVanni2011, Melas2014}.

The fragmentation algorithm is based on mapping
the initial agglomerate onto a symmetric adjacency matrix. Each
monomer configuration (\emph{i.e.,} cluster)
has a unique representation as a graph via the adjacency matrix.
A symmetric adjacency matrix $A_{ij}$ identifies completely a
non-directed graph. The adjacency matrix is used
to determine the connected components of a structure,
thence the fragments upon removal of a bond. A similar approach
was used by Isella and Drossinos~\cite{Lorenzo2010} to identify clusters generated
via monomer Langevin dynamics.

\begin{figure}[htb]
\onefigure[width=0.80\columnwidth,height=0.23\columnwidth]{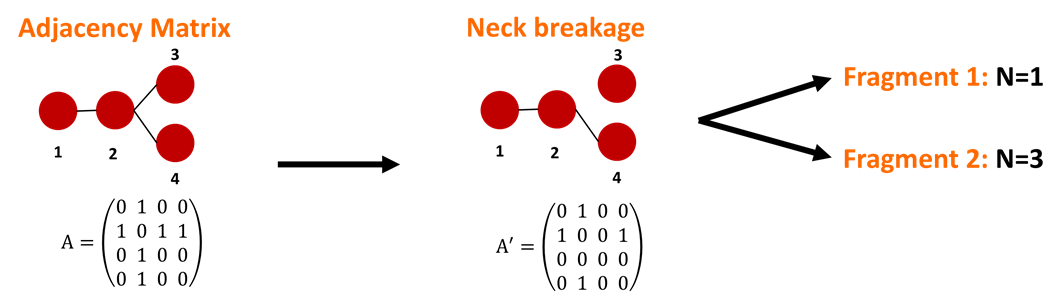}
\caption{Adjacency matrices associated with a fragmentation event (``neck breakage"). 
}
\label{fig:AdjacencyMatrixB}
\end{figure}

The adjacency matrix is constructed from all monomer-monomer Euclidean
distances: monomer-monomer links (bonds) are represented by one and their
absence by a zero.
Two monomers ($i,j$) are considered bonded if their
center-of-mass distance $D_{ij}$ is smaller than the monomer diameter $2 R_1$ plus
a threshold distance $D_{\textrm{thr}}$, i.e., $D_{ij} < 2 R_1 + D_{\textrm{thr}}$.
We took $D_{\textrm{thr}} = 10^{-3} \times 2R_1$.
A pictorial representation of a fragmentation event (referred to as ``neck breakage")
via adjacency matrices is shown in fig.~\ref{fig:AdjacencyMatrixB}.

The fragmentation algorithm consists of randomly choosing a non-zero element
of the adjacency matrix 
according to a discrete uniform probability distribution,
$\xi \, \epsilon \, U(0,N_{\textrm{link}})$,
where $N_{\textrm{link}}$ is the number of links in the agglomerate.
The selected element is replaced by a zero, thereby eliminating the monomer-monomer bond.
A new adjacency matrix is formed, and its connected components are determined.
If the initial agglomerate fragments, the sizes of the fragments are stored. If the aggregate does not fragment, it is removed from the list of initial clusters.
Only one link is removed per agglomerate, no multiple fragmentation events are considered. Then 
a new cluster is chosen, and the same procedure is repeated till the list
of clusters has been exhausted.
We found that fewer than 0.05\% of the clusters did not fragment. By removing them
we ensure that we consider only loop-less agglomerates.

\begin{figure}[htb]
\includegraphics[width=0.45\columnwidth,height=0.45\columnwidth]{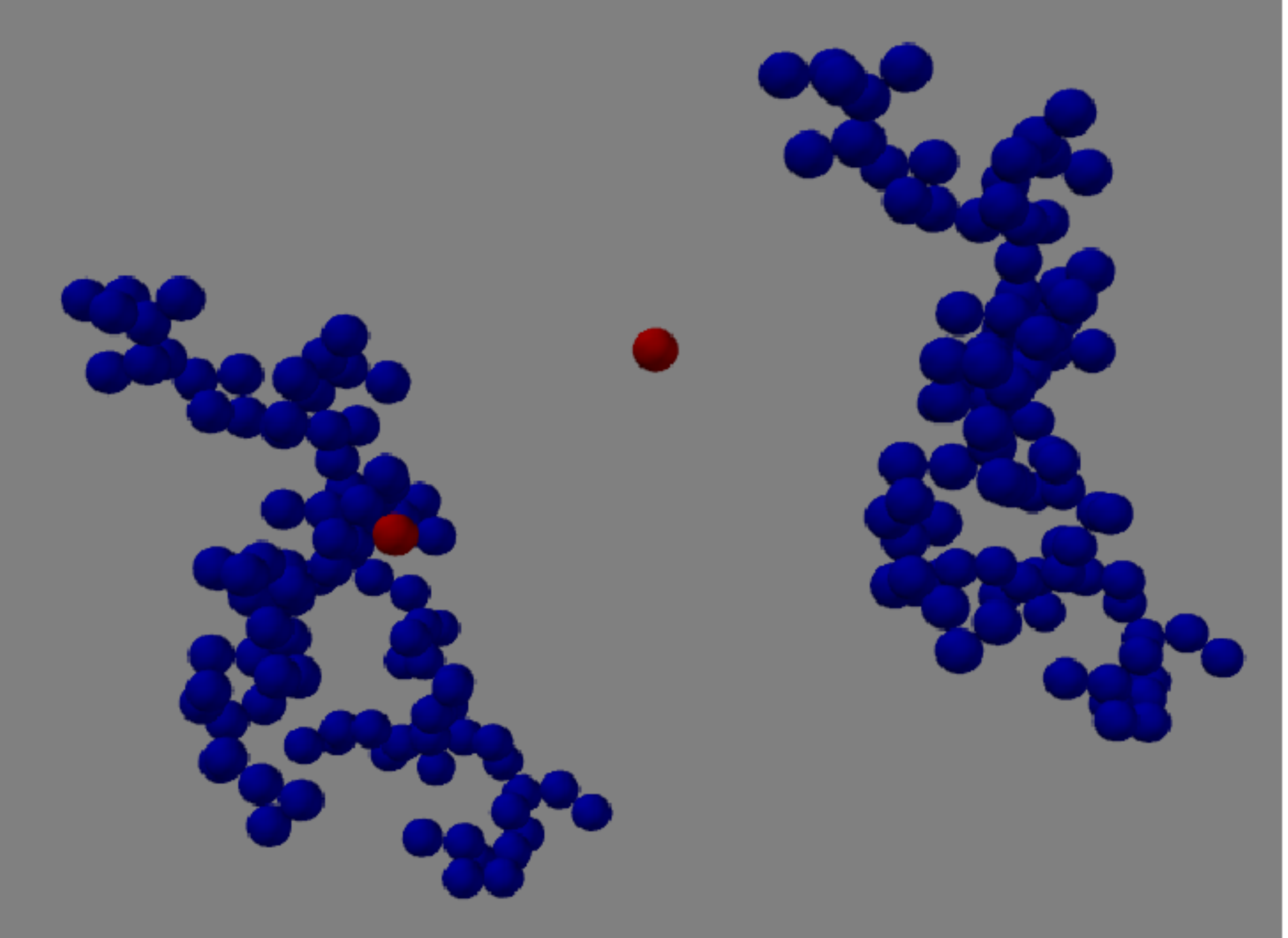}
\hspace{0.25cm}
\includegraphics[width=0.45\columnwidth,height=0.45\columnwidth]{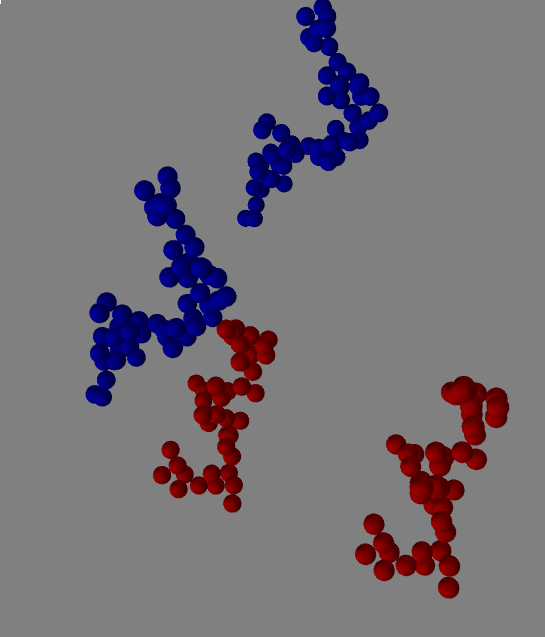}
\caption{Fragmentation events. On the left of each subfigure the initial cluster, on the 
right the resulting two fragments.
The red color denotes one of the fragments before and after fragmentation. Left:
Dissimilar fragments. Right: Equal-size fragments.}
\label{fig:FragmentationEvents}
\end{figure}

Figure~\ref{fig:FragmentationEvents} shows two simulated fragmentation
events. On the left subfigure the resulting fragments are dissimilar, on the right
they are of equal size. We found
that the fragments have a slightly smaller fractal dimension,
consequently their structure is more open, whereas they have 
a slightly higher fractal prefactor: they tend to be more locally compact than the 
initial agglomerate~\cite{Melas2014}. The structural parameters of the two fragments
were not identical: they showed a slight asymmetry, which disappears in the
fragment distribution.

\begin{figure}
\includegraphics[width=0.45\columnwidth,height=0.45\columnwidth]{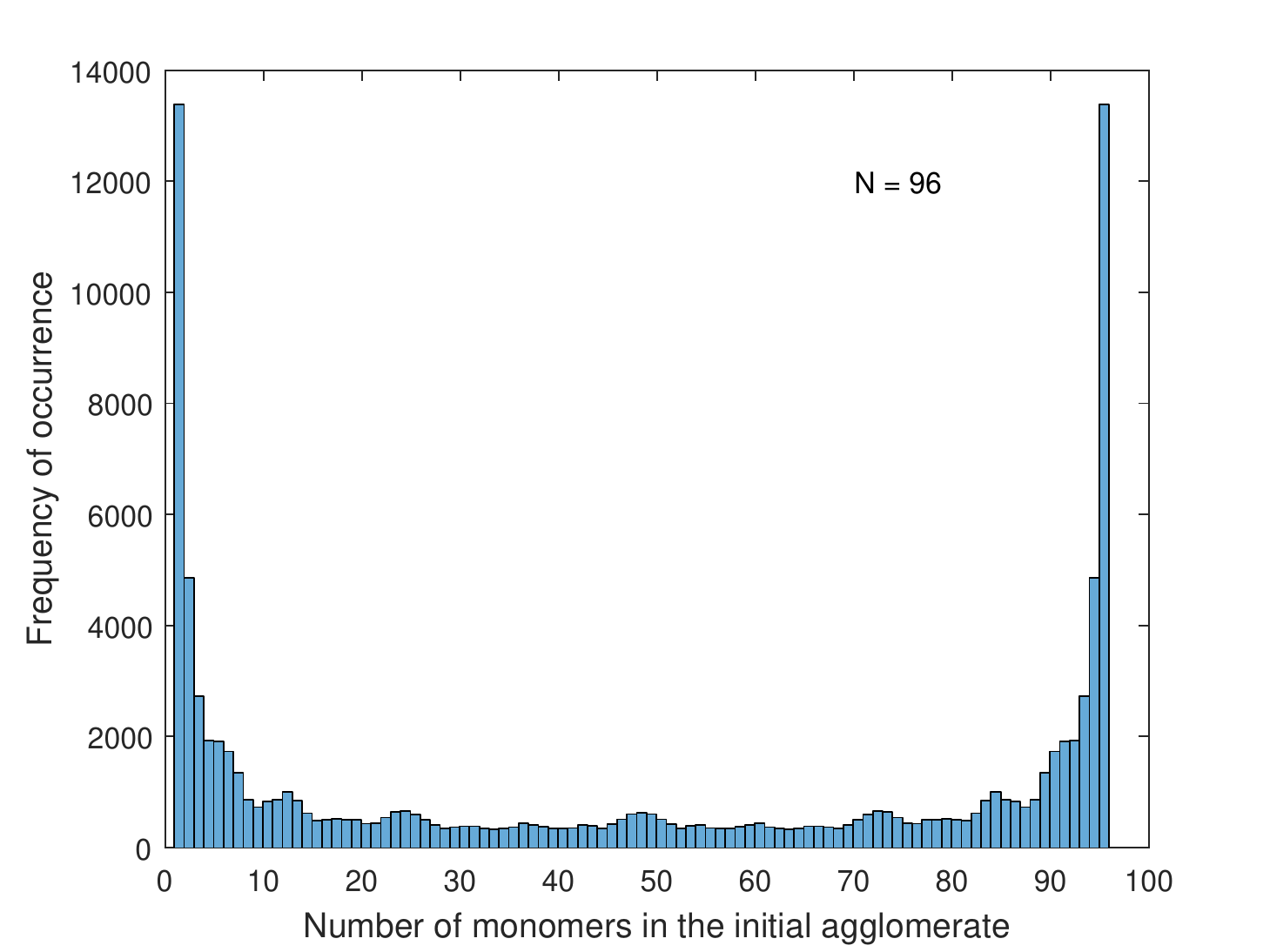}
\hspace{0.25cm}
\includegraphics[width=0.45\columnwidth,height=0.45\columnwidth]{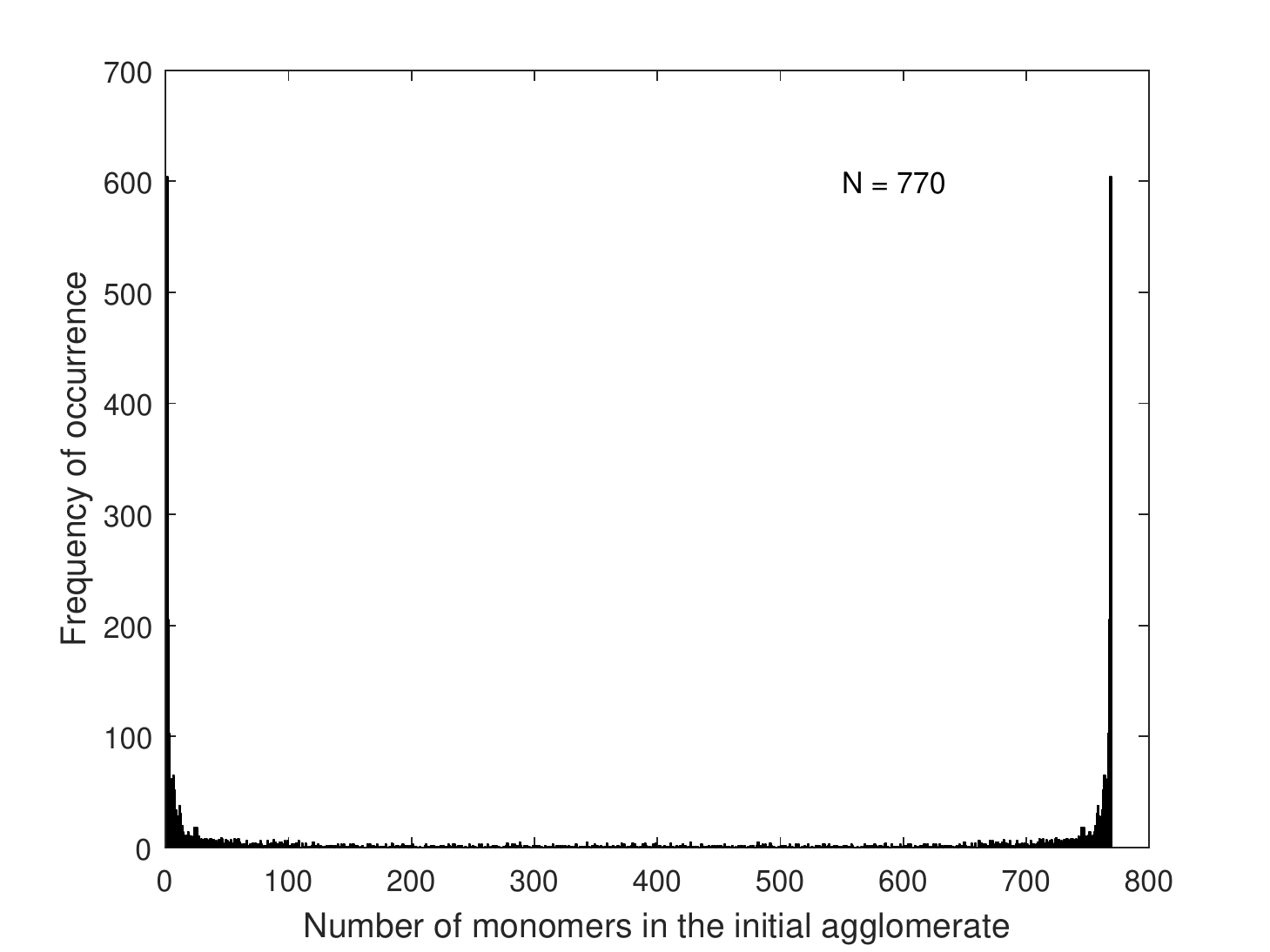}
\caption{Histograms of the fragments arising from the random, single-bond
fragmentation of DLCA agglomerates. Left: $j = 96$; Right: $j =770$.
The number of bins is $j-1$, the size of all possible fragments.}
\label{fig:Histograms}
\end{figure}

Figure~\ref{fig:Histograms} shows representative
histograms of the fragment size distribution upon single-bond random
fragmentation of small (left
subfigure, $j=96$) and large (right, $j=770$) Diffusion
Limited Cluster Aggregation (DLCA) agglomerates ($d_f = 1.8$, $k_f = 1.3$).
It is apparent that this fragmentation algorithm leads to
largely dissimilar fragment sizes: the resulting fragment size distribution is
U-shaped. Kalay and Ben-Naim~\cite{FragTrees2015} also found a U-shaped fragment
distribution in their study of the fragmentation of random trees by removing
a single node (a slightly different fragmentation process).
A predominance of non-equal-sized fragments upon bond removal
in random clusters on a lattice was also found in ref.~\cite{FragRandom2015}.
The experimental results of Kusters \emph{et al.}~\cite{Pratsinis1993}, who
found that ultrasonic fragmentation
of agglomerated particles in liquids leads to erosion, namely only a limited number
of monomers break up from the agglomerate, also suggest a U-shaped fragment size distribution.

\section{Empirical fragment size distribution}
\label{sec:FragSizeDistribution}

We fitted the empirically determined fragment distribution to
a symmetric beta distribution because it reproduces
a U-shaped distribution for certain values of its parameter.
Of course, our choice is not unique: other distributions could reproduce
our data, for example, eq.~(\ref{eq:Odriozola}). However, the beta-distribution
fit is more parsimonious: it depends on one parameter only
[for fixed number of monomers in the agglomerates and fixed aggregate morphology
($\df, \kf$)].
Moreover, a symmetric beta distribution was introduced \emph{ad hoc} previously
to model the fragment size probability density function~\cite{Ushape}, see
also ref.~\cite{McCoy1994}.
The functional form of the
fragment density function $p(x,y)$, expressed
in terms of the relative number of monomers $z= x/y$,
was chosen to be
\begin{subequations}
\beq
p(x,y; \df)  = \frac{2}{y} \, b[z; \beta(d_f, y)] , 
\eeq
with
\beq
b[z; \beta(d_f, y)] = \frac{[z(1-z)]^{\beta(d_f,y)}}
{\int_{\zmin}^{\zmax} \, \upd z \, [z(1-z)]^{\beta(d_f,y)}}
\eeq
for
\beq
\frac{1}{y} \equiv \zmin \leq
z \leq 1-\frac{1}{y} \equiv \zmax
\eeq
\label{eq:ContinuousFit}
\end{subequations}
By construction the continuous probability density function $b(z)$ is
symmetric, $b(z) = b(1-z)$, and it satisfies
the conservation laws 
\begin{subequations}
\beq
\int_{\zmin}^{\zmax} \upd z \, b[z; \beta(d_f,y)] = 1 , 
\eeq
\beq
\int_{\zmin}^{\zmax} \upd z \, z \, b[z; \beta(d_f,y)] = \frac{1}{2}.
\eeq
\label{eq:ConservationLawsCont}
\end{subequations}
Note that a fragment distribution
proportional to $[i^{-1}+(j-i)^{-1}]^\alpha$, a frequently used choice~\cite{ernst1987},
would lead to a beta-like dependence proportional to $[z(1-z)]^{-\alpha}$.
%

For a given initial cluster size $y$, and a fractal dimension, the fragment
distribution may be obtained by minimizing the distance,
in the sense of the $l^2$-norm, between the numerical results 
(the histograms shown in fig.~\ref{fig:Histograms} appropriately normalized)
and the predictions
of eqs.~(\ref{eq:ContinuousFit}).
The minimization procedure would minimize the distance
between $b(z) \upd z$ and $p_{ij}/2$ for all $i = 1, \ldots j-1$ to ensure
that $b(z) \upd z = p_{ij}/2$.
For $j-1$ fragments $\upd z = 1/j$, and the optimization condition becomes $b(z) = j p_{ij}/2$. This
procedure leads to a \emph{global} optimization, \emph{i.e.,} an optimization over
all fragment sizes.
%

\begin{figure}[htb]
\onefigure[width=0.75\columnwidth,height=0.5\columnwidth]{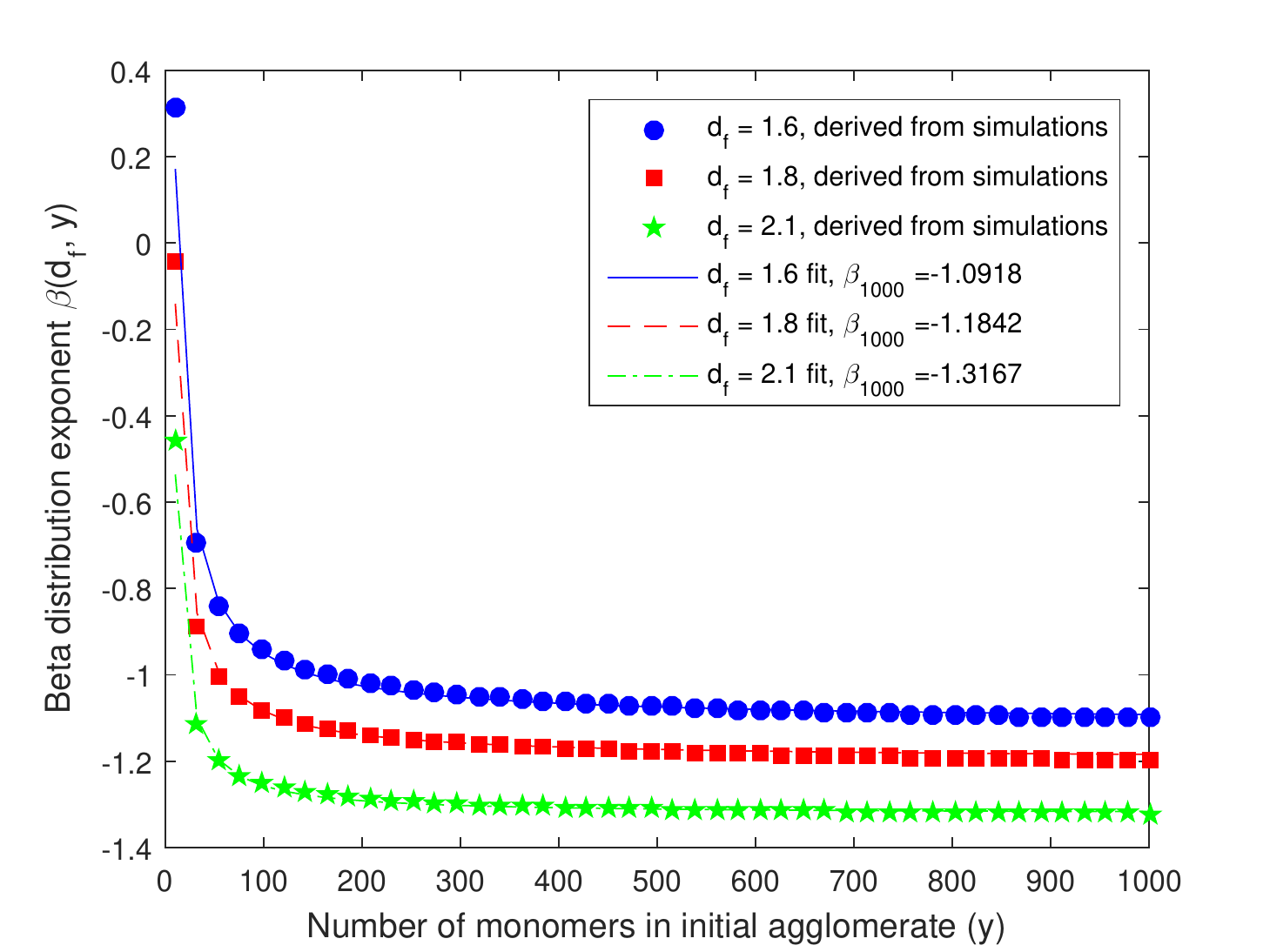}
\caption{Beta-distribution exponent $\beta(d_f,y)$ as a function of the initial agglomerate size,
parametrized by the fractal dimension. 
Numerical simulations (filled symbols) are compared to
calculations via eqs.~(\ref{eq:BetaContinuousBoth}) (lines);
$k_f = 1.3$.}
\label{fig:ExponentBetaCont}
\end{figure}

Instead, we opted for a
\emph{local} optimization.
It is more
important to reproduce the small-fragment behaviour ($z \to 1/y$ or $z \to 1-1/y$),
the size of most fragments, 
than the complete distribution. Moreover, since the number of
equal-size fragments is relatively small, the quality of the simulation data
deteriorates as $z \to 1/2$. We determined the probability of occurrence of
a monomer fragment for all
the clusters we generated 
for the three fractal dimensions. We found that
\beq
g_{1y} = \frac{1}{2} \, p_{1y} = -0.1442 + 0.1518 d_f , \quad \forall y,
\label{eq:FitMon}
\eeq
provides an excellent fit of our simulation data.
Equation~(\ref{eq:FitMon}) suggests that $p_{1y}$ depends
only on the fractal dimension and not on the size of the initial agglomerate.
The exponent $\beta$ was obtained by requiring that the 
probability of monomer occurrence
\beq
g_{1y} =  \frac{1}{y} \, b[\frac{1}{y}; \beta(d_f,y)]
= \frac{1}{y} \, \frac{[\frac{1}{y}(1-\frac{1}{y})]^{\beta(d_f,y)}}
{\int_{\zmin}^{\zmax} \, \upd z \, [z(1-z)]^{\beta(d_f,y)}} , 
\label{eq:MonomerFit}
\eeq
equal the empirically determined monomer occurrence probability.
We used eq.~(\ref{eq:FitMon}) to generate monomer occurrence probabilities for
$j =10, 11, \ldots , 1010$, and we fitted the resulting beta-distribution exponent to
\begin{subequations}
\beq
\beta(d_f,y) = a(d_f) + b(d_f) \, y^{c(d_f)}.
\label{eq:BetaFit}
\eeq
to obtain the coefficients
\begin{align}
a(d_f) & = -0.42 \, d_f - 0.44;
\quad b(d_f) = -1.28 \, d_f + 12.29; \nonumber \\
c(d_f) & = -0.37 \, d_f - 0.31, \quad \mbox{\textrm{for}} 
\ \df \, \epsilon \, [1.6, 2.1].
\label{eq:BetaCoefficients}
\end{align}
\label{eq:BetaContinuousBoth}
\end{subequations}
We show the exponent $\beta(d_f,y)$ in fig.~\ref{fig:ExponentBetaCont}.
It tends to a constant asymptotic value as $y \to \infty$.
In that limit the
fragment density function $p(x,y; \df)$ becomes approximately
a homogeneous function
under the scaling $(x,y) \rightarrow (\lambda x, \lambda y)$ 
since $\lim_{y \to \infty} \beta(\df,y) = \beta(\df)$.
%
The kernel does not become exactly homogeneous because
some $y$-dependence remains in the integration limits. 
The $y$-dependent limits are
required because
the asymptotic exponent is $\beta(d_f, \infty) < -1$ (for the studied fractal dimensions).
The beta distribution is properly defined only for $\beta > -1$, the
reason why our empirical fit eq.~(\ref{eq:ContinuousFit})
is defined for $z \, \epsilon \, (0,1)$.
This is not a limitation
since there is always a finite monomer size~\cite{McCoy1994}.
Homogeneous fragmentation
kernels have been used extensively in the past, see, \emph{e.g.,}
refs.~\cite{McGradyZiff1987,VigilZiff1989,Kostoglou1997,Redner}.
We found that the coefficients presented in eq.~(\ref{eq:BetaCoefficients})
depend slightly on whether a discrete or
continuous distribution was modelled. The differences
are small, but noticeable; see fig.~\ref{fig:CompareBetaDiscreteCont}
in Appendix C (Supplementary Material).

\begin{figure}[htb]
\onefigure[scale=0.45]{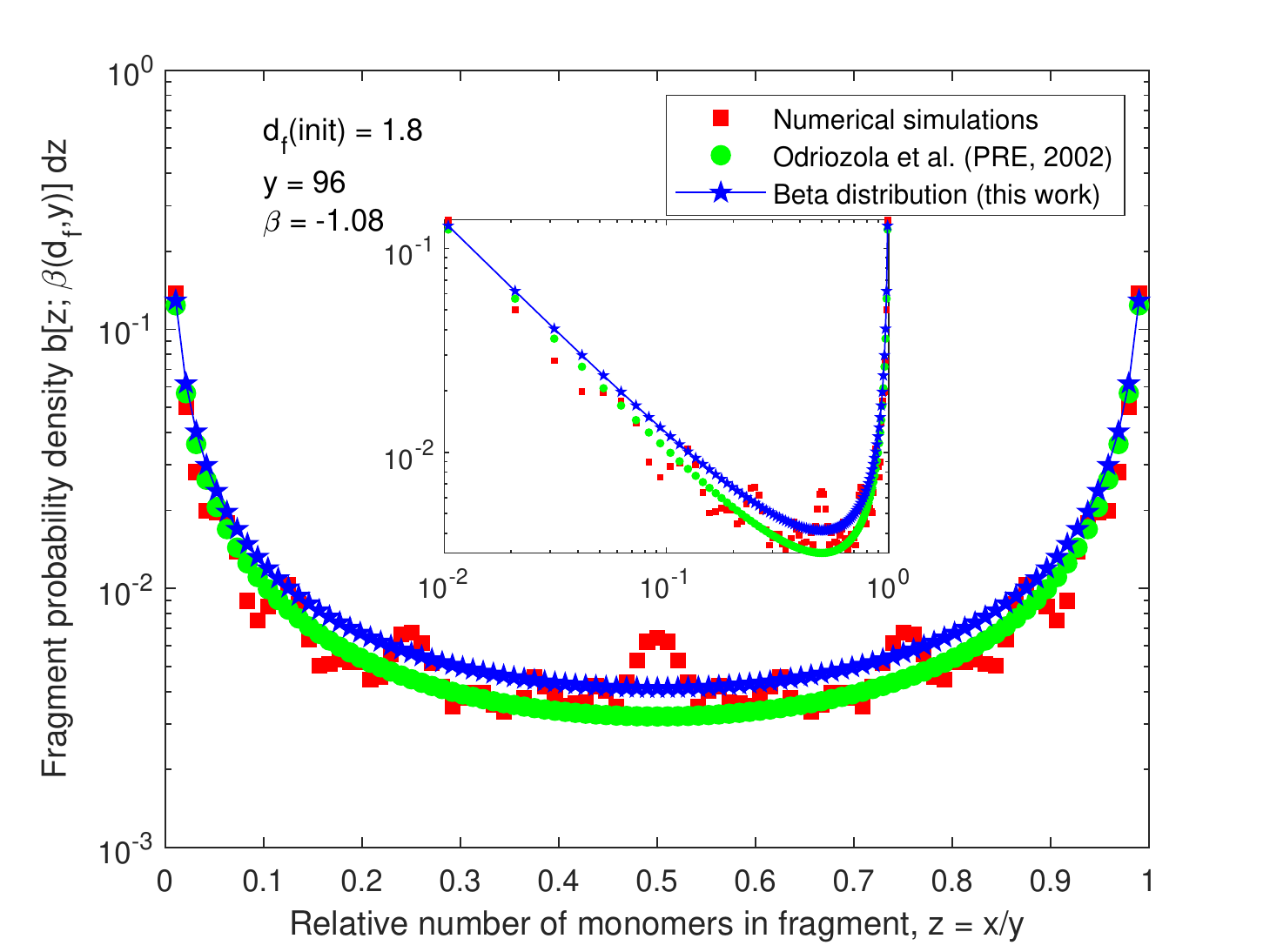}
\onefigure[scale=0.45]{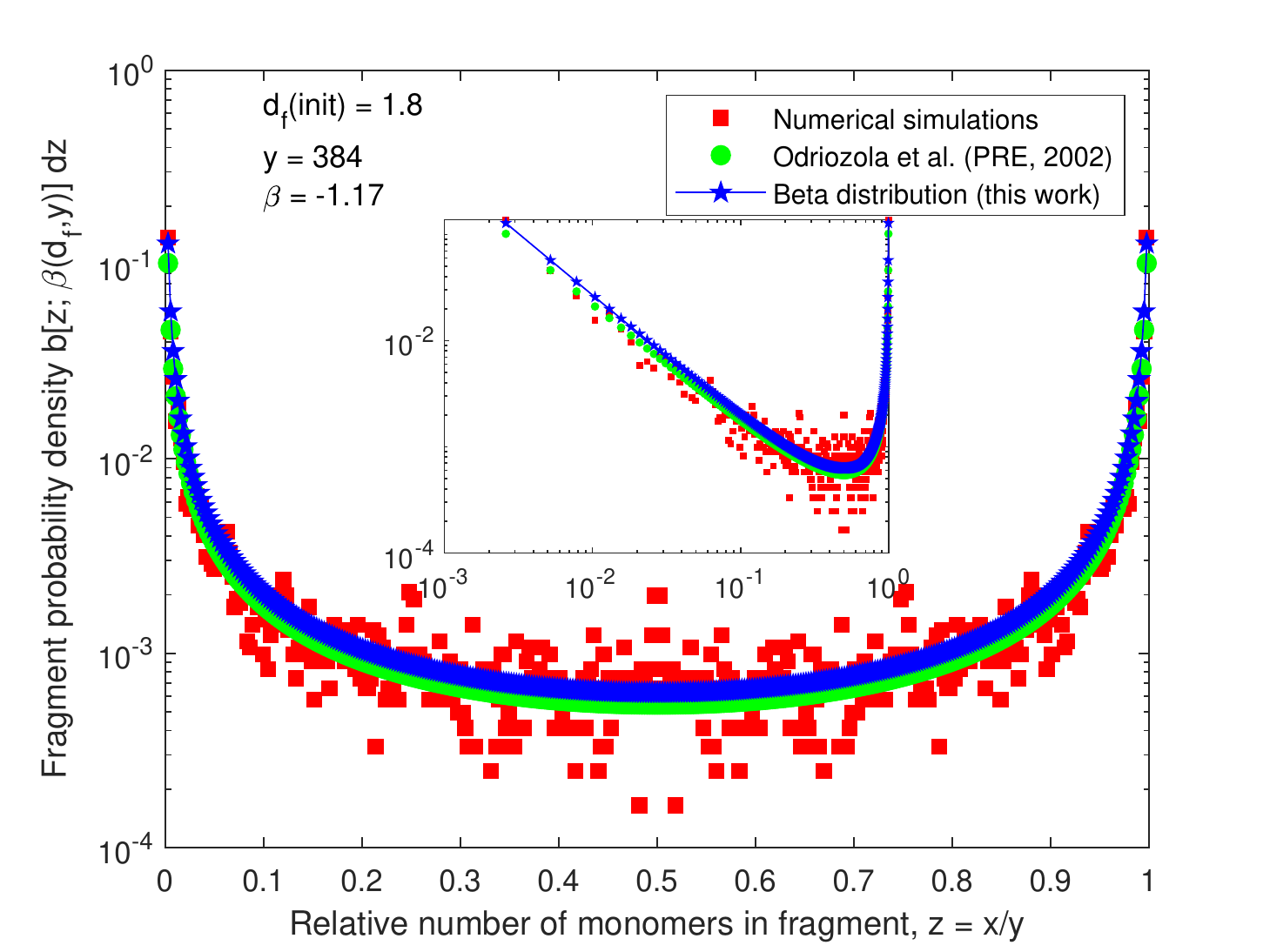}
\caption{DLCA random fragmentation. Numerical fragment size 
probability distribution
(red square symbols); ref.~\cite{Odriozola2002}, eq.~(\ref{eq:OdriozolaPlot}) (green circles);
beta distribution ($b[z;\beta(\df,y)] \upd z; \upd z = y^{-1}$,
blue pentagrams).
Top: $y = 96$; Bottom: $y = 384$ monomers. Inset: double logarithmic plot.}
\label{fig:FullDist18}
\end{figure}

Predictions of our empirical fit are compared to numerical simulations of 
single-bond, random fragmentation of DLCA
agglomerates in fig.~\ref{fig:FullDist18}.
Simulation results
(filled square red symbols) were obtained by normalizing (to unity) the histogram
of fragment sizes with $j-1$ bins. The beta distribution points,
converted from the continuous density function to a discrete mass function,
are denoted by filled blue pentagrams
(note the logarithmic y-scale of the main figure; inset double logarithmic plot).

We also compare the numerical distributions to an expression
proposed by
Odriozola et al.~\cite{Odriozola2002}, who
analysed the fragment distribution of DLCA clusters generated by a
slightly different algorithm. They
fitted the fragment size distribution, specifically, $f_{ij}$ which is
the number of bonds that upon breaking would divide a cluster into
fragments of sizes $i$ and $j-i$, to a four-parameter function, 
\beq
f_{ij} = p_1 \, \big [ i^{p_2} + (j-i)^{p_2} \big ] \,
\big [ i^{-p_3} + (j - i)^{-p_3} \big ] \, 
\big [ i (j-i) \big ]^{p_4} ,
\label{eq:Odriozola}
\eeq
with $p_1 = 0.4391$, $p_2 = 1.006$, $p_3 = 1.007$, and $p_4 = -0.1363$. 
The product of the second term times the third term
is reminiscent of the continuum Brownian
agglomeration kernel for two fractal-like clusters of dimension $d_f$.
Note that $p_2 \sim p_3 \sim 1$, suggesting that eq.~(\ref{eq:Odriozola}) approximates
a symmetric
beta distribution with a fixed exponent, independent of $j$, and for a fixed fractal dimension
($d_f = 1.8$).

\begin{figure}[htb]
\begin{center}
\onefigure[scale=0.46]{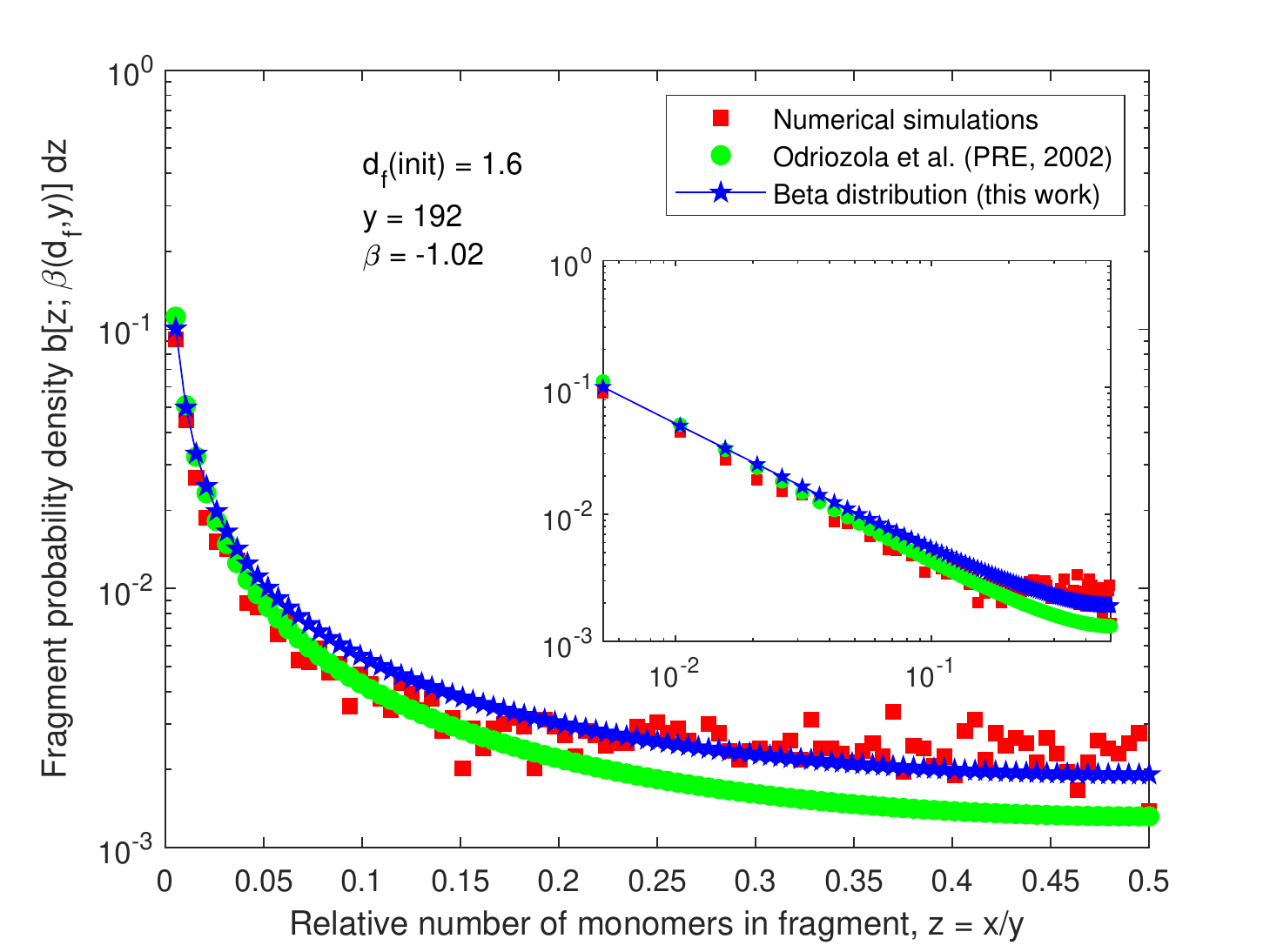}
\onefigure[scale=0.46]{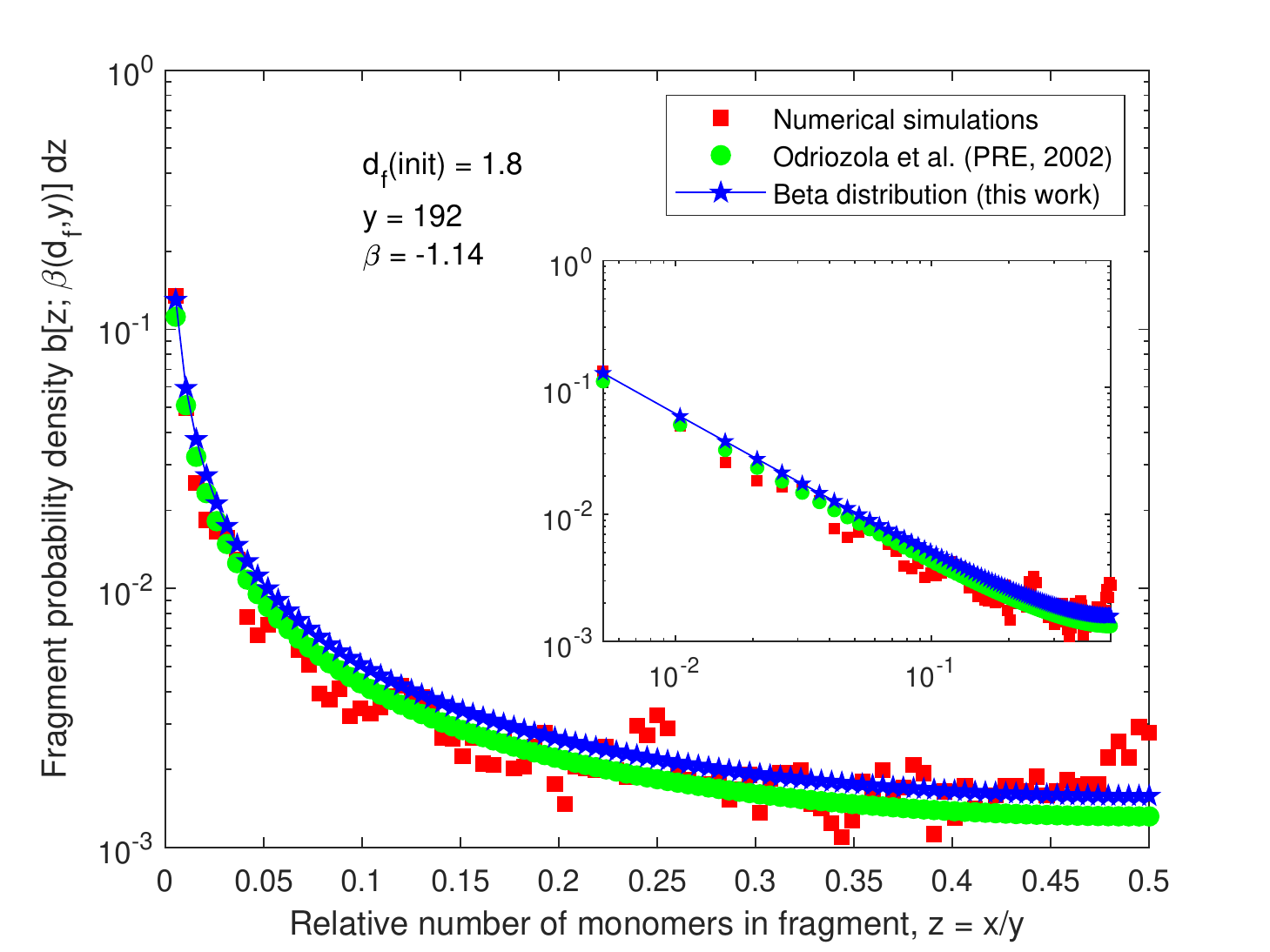}
\onefigure[scale=0.46]{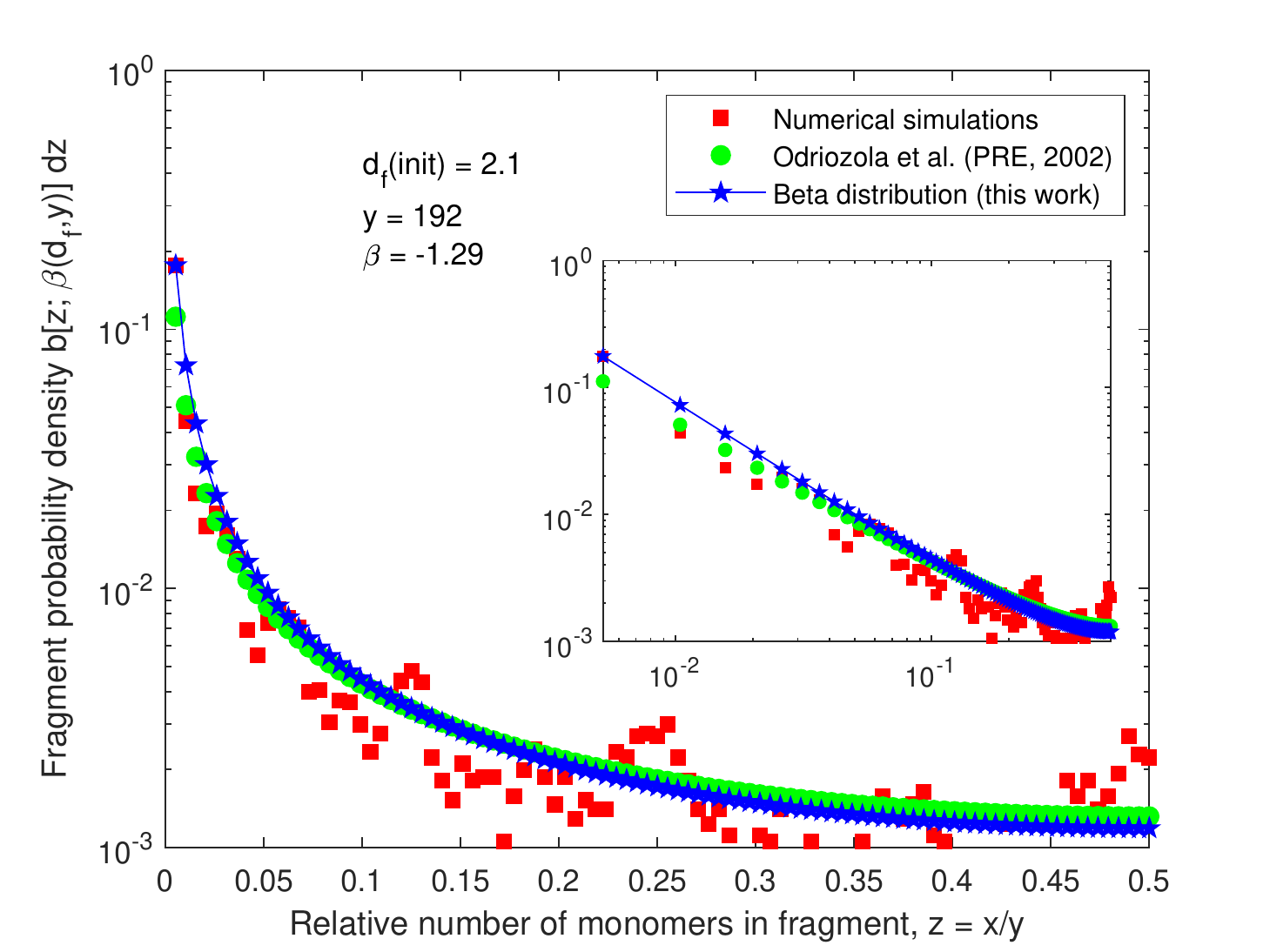}
\caption{Morphology-dependent fragment size distribution: numerical simulations
(red square symbols), ref.~\cite{Odriozola2002} (green circles), and
beta distribution ($b[z;\beta(\df,y)] \upd z; \upd z = y^{-1}$, blue pentagrams).
Top $d_f = 1.6$; middle $d_f = 1.8$, bottom
$d_f = 2.1$; $y = 192$, $k_f = 1.3$. Inset: Double logarithmic plot.}
\label{fig:FullDist}
\end{center}
\end{figure}

\begin{figure}[htb]
\begin{center}
\onefigure[scale=0.46]{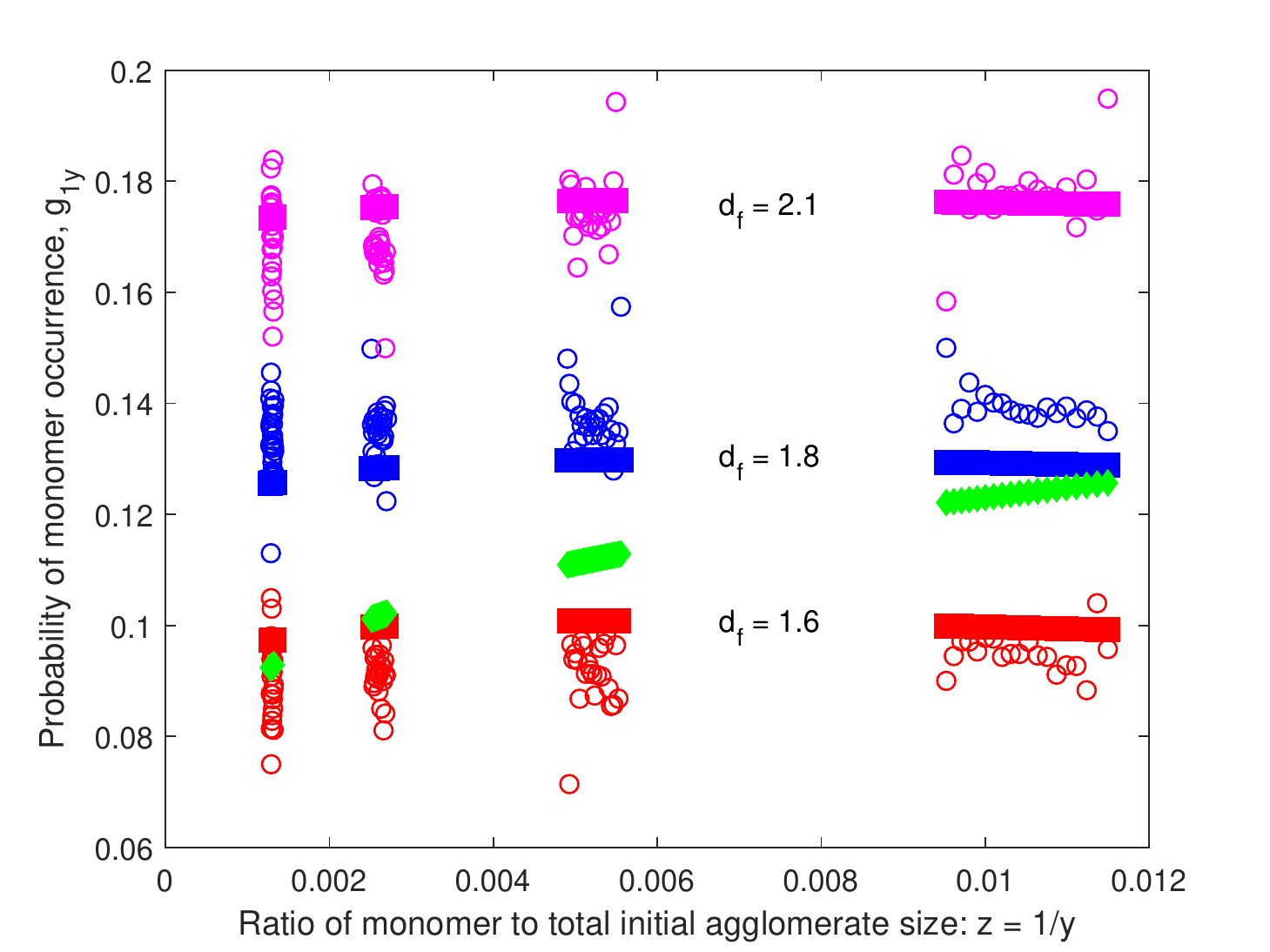}
\caption{Probability of occurrence of a monomer (or $y-1$) fragment as a function of
the initial-agglomerate size, parametrized by its fractal dimension. Open symbols denote simulations, filled symbols theoretical
predictions. Magenta, blue, and red colors
denote this work, $b[y^{-1}, \beta(d_f,y)]/y$, eqs.~(\ref{eq:MonomerFit},
\ref{eq:BetaContinuousBoth}); Green squares denote eq.~(\ref{eq:OdriozolaPlot}).}
\label{fig:ProbMonOccurrence}
\end{center}
\end{figure}

As argued earlier, eq.~(\ref{eq:Odriozola}) may by normalized to unity by
dividing it by the total number of bonds to obtain
\beq
g_{ij} = \frac{f_{ij}}{2 (j-1)} .
\label{eq:OdriozolaPlot}
\eeq
The predictions of eq.~(\ref{eq:OdriozolaPlot}) are denoted by filled, green circles
in figs.~\ref{fig:FullDist18}, \ref{fig:FullDist}, and \ref{fig:ProbMonOccurrence}.
We note reasonable agreement of simulation data with theoretical predictions.
It should be stressed that the normalization condition $\sum_{i=1}^{j-1} f_{ij}$ is
\emph{approximately} equal to $2(j-1)$. In fact, the authors
of ref.~\cite{Odriozola2002} used this condition to
estimate the accuracy of their fit. This implies  that 
eq.~(\ref{eq:OdriozolaPlot}) does not ensure mass conservation,
rendering its use in population balance
equations problematic. This is in sharp contrast to eqs.~(\ref{eq:ContinuousFit}) that
by construction satisfy exactly mass conservation.

The effect of morphology is studied in fig.~\ref{fig:FullDist}.
The two empirical fits are compared graphically
to numerically determined fragment distributions arising from the
break up of $192$-monomer clusters for the three
fractal dimensions. 
As the fractal dimension increases the probability
of obtaining small fragments increases, the fragment
distribution sharpens. As before the agreement is very good,
even though eq.~(\ref{eq:Odriozola}) was derived for relatively
small DLCA clusters ($j<100$).

We also present the probability of occurrence of a monomer fragment as a function
of initial agglomerate size parametrized by the fractal dimension (of the initial agglomerate)
in fig.~\ref{fig:ProbMonOccurrence}. Since the exponent $\beta$ was determined from
the monomer probability, our proposed fit reproduces very well the simulation data,
and it shows a dependence on the fractal dimension.
As the fractal dimension decreases the probability of monomer occurrence decreases,
In fact, we expect that as the fractal dimension decreases the
fragment distribution would flatten out:
the straight-chain limit ($\df = 1$)
is a constant probability distribution
independent of fragment size, but dependent on $y$, 
$b(z;\df=1) =1, p(x,y;1) = 2/y$.

\section{Universal fragment size probability distribution}
\label{sec:UniversalFragDistribution}

\begin{figure}[htb]
\onefigure[scale=0.45]{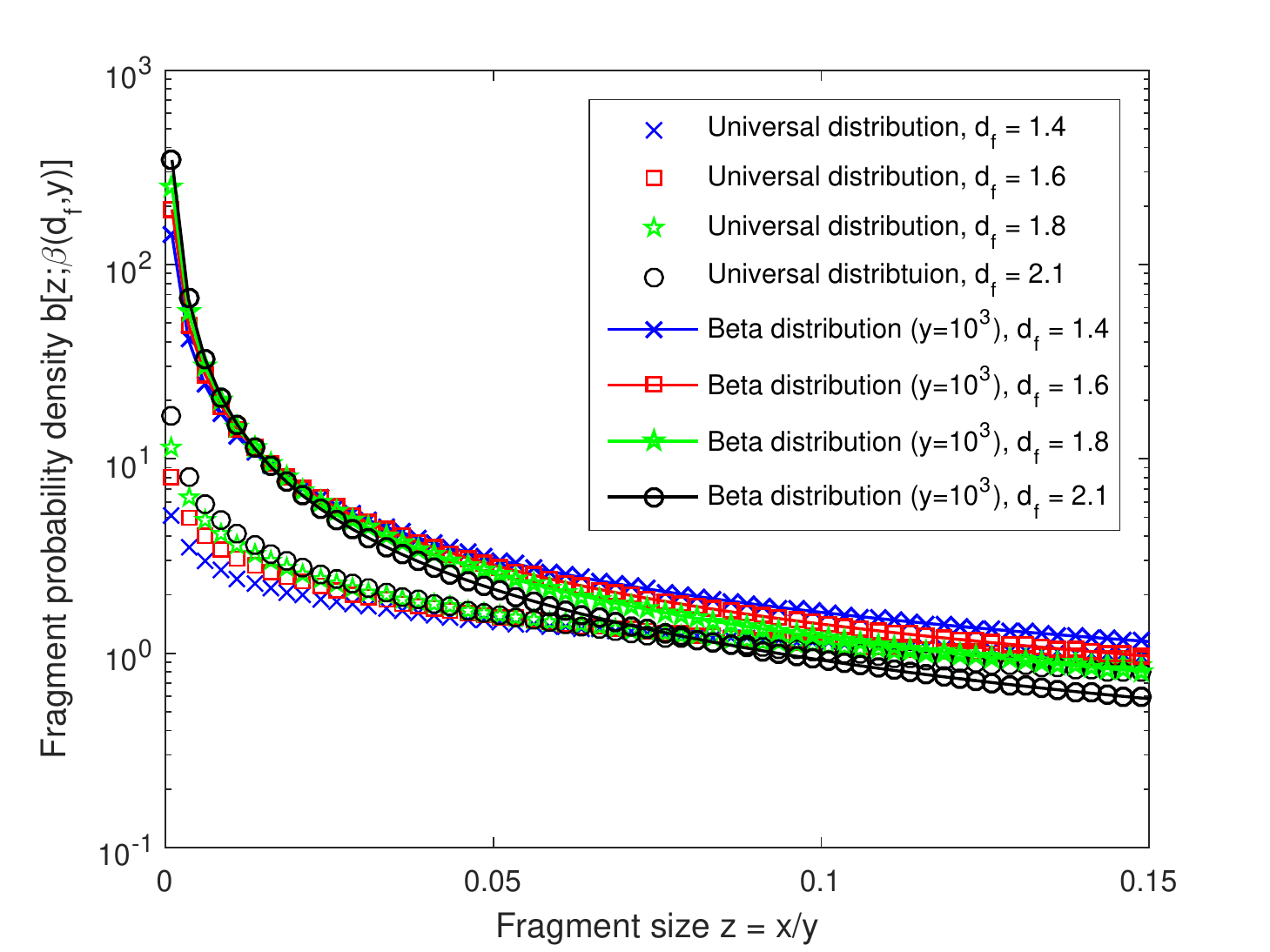}
\onefigure[scale=0.45]{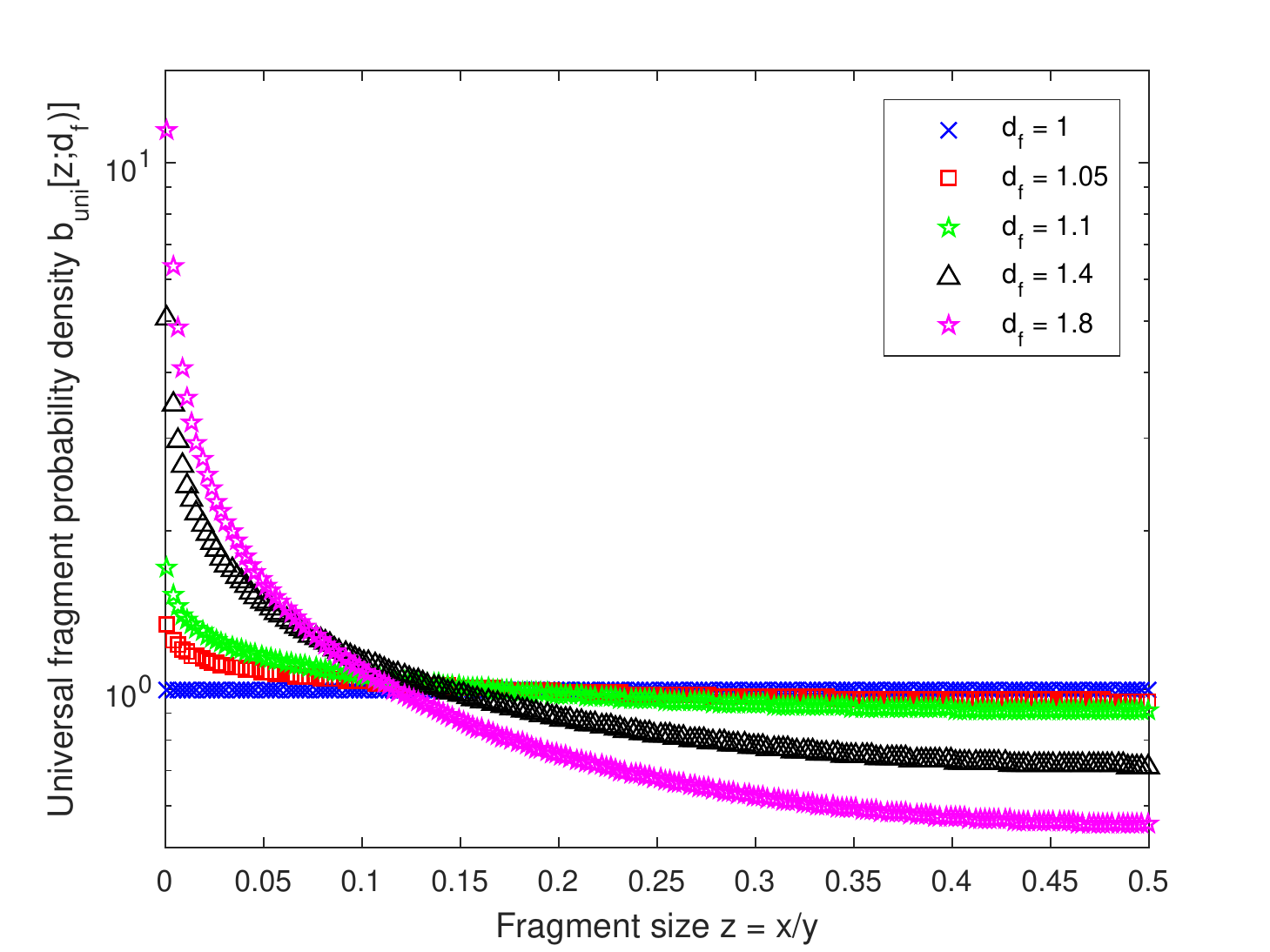}
\caption{Universal fragment size distribution eq.~(\ref{eq:NoParameters}). Top: Comparison to
the beta distribution at the asymptotic
exponent $\beta(\df, 1000)$; Bottom: Approach
to the straight-chain, uniform, limit ($d_f = 1$).}
\label{fig:UniFragDist}
\end{figure}

We also studied the effect of morphology through a universal fragment size
probability density function (a parameter-free distribution) that
depends only on the initial cluster fractal dimension. We constructed it
by requiring that it satisfy exactly
the conservation laws eqs.~(\ref{eq:ConservationLawsCont}) and that it have the correct
limit for straight chains. 
The universal fragment size probability distribution is
\begin{widetext}
\begin{equation}
\label{eq:NoParameters}
b_{\textrm{uni}}(z; d_f) =
\Big  \{ d_f \, z(1-z) \, [ z^{1/d_f} + (1-z)^{1/d_f} ] \,  [ z^{-1/d_f} + (1-z)^{-1/d_f} ]
\Big \}^{-1} . 
\end{equation}
\end{widetext}
\begin{floatequation}
\mbox{\textit{see Eq.~\eqref{eq:NoParameters}}} \nonumber
\end{floatequation}
It is easy to show that the conservation law is satisfied:
$\int_0^1 \upd z\, b_{\textrm{uni}}(z;d_f ) = 1$ as calculated
analytically by Mathematica~\cite{Mathematica}.
In the straight-chain limit 
$b_{\textrm{uni}}(z, \df = 1) = 1$ ($p(x,y; \df = 1) = 2/y$). 
Moreover, since eq.~(\ref{eq:NoParameters}) is a function of $z$ only and independent of $y$,
under the 
scaling of sizes, $b_{\textrm{uni}}(z;d_f)$ remains invariant. Its homogeneity
index is zero, leading to a homogeneous fragmentation kernel.
The predictions of eq.~(\ref{eq:NoParameters})
are compared to those of the beta distribution in
fig.~\ref{fig:UniFragDist}, top subfigure [for the asymptotic
beta-distribution exponent $\beta(\df,1000)$]. It is apparent
that the parameter-free distribution does not reproduce
the simulation data for $d_f \neq 1$ as accurately as the beta distribution.
However, it does reproduce the straight-chain limit, bottom subfigure.
That subfigure shows that the effect of the fractal dimension becomes
significant only for fractal dimensions very close to the singular limit $d_f \to 1$.
The transition from a pronounced
U-shaped distribution to the limiting case 
of a constant (flat) distribution of a straight line becomes noticeable for
$d_f < 1.1$ in that at $d_f=1.1$ the probability of a monomer fragment occurring is still
predicted to be almost twice that of a monomer fragment of a straight chain. 

\section{Conclusions}
\label{sec:Conclusions}

We studied numerically the fragment size distribution upon random scission
of \emph{in silico} fractal-like agglomerates as a function of the initial size of the
agglomerate and its structural properties, 
fractal dimension
($\df = 1.6, 1.8, 2.1$) and fixed prefactor ($\kf = 1.3$).
The fragment size distribution is an
essential ingredient of the fragmentation kernel.
Earlier studies, for
example refs.~\cite{Meakin1988,Kostoglou1997,ernst1987,Redner},
used fragmentation kernels in population
balance equations whose analytical form was
dictated either by physical arguments
or by properties of related homogeneous agglomeration kernels,
instead of originating from simulation
data or experimental measurements.
We followed the latter approach to obtain an analytical
expression for
the fragment size distribution
that reproduces
rather accurately  our numerical simulations. 

Our numerical simulations showed that,
for our fragmentation algorithm, the fragment size distributions are U-shaped, namely most
clusters fragment into largely dissimilar fragments. We showed that
a symmetric beta distribution reproduces rather accurately the empirical
fragment size distributions. 
We found that the beta-distribution exponent 
depends on the initial agglomerate morphology (via the fractal dimension for
fixed prefactor) and the number of monomers $y$
in the initial agglomerate, tending to a $\df$-dependent
asymptotic limit for large $y$.

We, also, derived a universal (parameter-free) fragment size distribution,
dependent on the fractal dimension of the initial cluster,
by requiring that it satisfy the two fragmentation
conservation laws 
and that it reproduce the straight-chain limit. 



\acknowledgments
The views expressed are purely those of the authors and may not in any circumstances be
regarded as stating an official position of the European Commission. A.D. Melas was
partially supported
by the Horizon 2020 EU Framework Programme through the SUREAL-23 project (Grant Agreement 724136).

\bigskip

\begin{center}
\textbf{Supplementary Material}
\end{center}


\section{Appendix A: Fragmentation conservation laws}

We first show that the mass conservation law, eq.~(3) in the main text, for the
discrete fragment size distribution is
a direct consequence of the first conservation law [expected number of fragments, eq.~(2)].
The derivation is based on splitting the sum into two parts,
redefining a summation index, and then using the symmetry property of the
fragment size distribution, $p_{ij} = p_{(j-i)j}$. Specifically,
\begin{align}
\sum_{i=1}^{j-1} i \, p_{ij} & = \sum_{i=1}^{(j-1)/2} i \, p_{ij} 
+ \sum_{i=(j-1)/2}^{j-1} i \, p_{ij} \nonumber \\ 
& \mbox{\textrm{(decompose the sum)}} \label{eq:Mass1} \\
&  = \sum_{i=1}^{(j-1)/2} i \, p_{ij}+ \sum_{i=1}^{(j+1)/2} (j-i) \, p_{(j-i)j} \nonumber \\ 
& \mbox{\textrm{(change summation index, second sum)}} \label{eq:Mass2} \\
&  = \sum_{i=1}^{(j-1)/2} i \, p_{ij}+ \sum_{i=1}^{(j-1)/2} (j-i) \, p_{(j-i)j} \nonumber \\ 
& + \sum_{i=(j-1)/2}^{(j+1)/2} (j-i) \, p_{(j-i)j} \nonumber \\ 
& \mbox{\textrm{(split second sum)}} \label{eq:Mass3} \\
&  = \sum_{i=1}^{(j-1)/2} i \, p_{ij}+ \sum_{i=1}^{(j-1)/2} (j-i) \, p_{ij} \nonumber \\
& + \sum_{i=(j-1)/2}^{(j+1)/2} (j-i) \, p_{ij} \nonumber \\ 
& \mbox{\textrm{(use $p_{ij}$ symmetry property)}} \label{eq:Mass4} \\
&  = j \sum_{i=1}^{(j-1)/2} p_{ij} + \sum_{i=(j-1)/2}^{(j+1)/2} (j-i) \, p_{ij} . \label{eq:Mass5}
\end{align}

A similar decomposition of the first conservation law leads to
\beq
\sum_{i=1}^{(j-1)/2} p_{ij} = 1 - \frac{1}{2} \, \sum_{i=(j-1)/2}^{(j+1)/2} p_{ij}
\label{eq:Frag1} .
\eeq
Substitution of eq.~(\ref{eq:Frag1}) into eq.~(\ref{eq:Mass5}) gives
\beq
\sum_{i=1}^{j-1} i \, p_{ij} = j + \sum_{i=(j-1)/2}^{(j+1)/2} (\frac{j}{2}-i) \, p_{ij}  = j .
\label{eq:AlmostFinal}
\eeq
The second sum in eq.~(\ref{eq:AlmostFinal}) is zero because the factor
$j/2-i$ changes sign with respect to $j/2$, whereas the distribution $p_{ij}$ is symmetric.
Specifically for an odd number of monomers in the cluster $j = 2n+1$ the sum becomes
\beq
\sum_{i=n}^{n+1} (n + \frac{1}{2}-i) \, p_{i(2n+1)} = \frac{1}{2} \, p_{n(2n+1)} -
\frac{1}{2} \, p_{(n+1)(2n+1)} = 0,
\label{eq:OddNumber}
\eeq
since $p_{(n+1)(2n+1)} = p_{n(2n+1)}$.
For $j = 2n$ even, the relevant fragment sizes
are $n - (1/2)$ and $n+(1/2)$. Since they are non-integers
the distribution functions $p_{(n-(1/2))(2n)}$ and $p_{(n+(1/2))(2n)}$
are zero as fragments do not have non-integer number of monomers (alternatively, if half-integer
fragments are considered, the previous
argument for the sum being zero due to the symmetry properties of $p_{ij}$ still holds ensuring
that the sum is zero even for an even number of monomers in the initial cluster).
Hence, for the discrete fragment distribution, the first conservation law (expected
number of fragments) along with the symmetry property of the fragment size
distribution imply the second conservation law (mass conservation),

The derivation applies \emph{mutatis mutandis} to the continuous distribution $p(x,y; d_f)$.
Specifically, the mass conservation law becomes
\beq
\int_1^{y-1} \mathrm{d} x \, x \, p(x,y) = y + \int_{(y-1)/2}^{(y+1)/2} \mathrm{d}x \,
\Big ( \frac{y}{2} - x \Big ) \, p(x,y)  = y .
\label{eq:Cont}
\eeq
The integral on the RHS of eq.~(\ref{eq:Cont}) vanishes because the
function $( y/2 - x ) p(x,y)$ is an odd function with respect to $y/2$.
Therefore, as in the case of the discrete fragment size distribution,
only one conservation law is independent.

\section{Appendix B: Cluster generation and description}
\label{sec:AppendixA}

The tunable cluster-cluster agglomeration
algorithm~\cite{ThouyJullienSI, FilippovAlgoSI} used to generate
the synthetic agglomerates 
is hierarchical in that after the generation of the initial
building blocks, the clusters combine pairwise.
A new cluster is generated from two pre-existing
clusters by first choosing randomly
a sticking point and a sticking angle: the two clusters stick at that
point with the appropriate orientation. The two random choices ensure that the resulting agglomerate
is unique. Then, one of the initial clusters is rotated by randomly choosing the
three Euler angles. A distance condition on the rotated cluster, which ensures that the resulting
cluster would satisfy the scaling law, is checked. If satisfied, the code
checks whether monomers overlap: if they do not, the new
agglomerate is accepted~\cite{Melas2014SI}.

Each pairwise binding 
defines a generation. We considered two types of initial block units:
dimers ($N_{\textrm{init}} = 2$) and a collection of $k$-mers
randomly chosen to have between six and eight monomers,
$N_{\textrm{init}} = 6,7,8$. The number
of monomers in a cluster of generation $n$ is $N_{\textrm{init}} \times 2^n$.
The random choice of 
initial building blocks of a varying number of monomers gives
clusters containing a range of monomer numbers, centered about a mode
(the most frequently occurring number of monomers in a set of clusters)
that depends on the number of monomers in the initial building blocks (and the
generation number).
We tested two slightly different
generation algorithms because we noted that random fragmentation
of clusters generated only with dimers as the initial blocks
tended to produce fragments containing
a ``magic" number of monomers (usually multiples of 2). Hence,
fragmentation is a severe test of the cluster-generation algorithm.
All the clusters considered in this work were generated with initial blocks
of varying number of monomers. 

\begin{largetable}
\caption{Total number, size range (the minimum and maximum number of
monomers in the set of $n$-th generation clusters), and mode (the most
frequently occurring number of monomers in the set of $n$-th generation clusters)
of the synthetic agglomerates ($k_f = 1.3$).}
\begin{center}
\label{table:Agglomerates}
\begin{tabular}{cccccc} \hline \hline
$d_f$ & Gen 4 & Gen 5 & Gen 6 & Gen 7 & Total number \\
& Size range & Size range & Size range & Size range & \\
& Mode & Mode & Mode & Mode & \\ \hline
1.6 & 200,000 & 100,000 & 50,000 & 25,000 & 375,000 \\
& 82 to 109 & 173 to 210 & 355 to 410 & 729 to 806 & \\
& 96 & 192 & 385 & 769 & \\
1.8 & 400,000 & 200,000 & 100,000 & 50,000 & 750,000 \\
& 83 to 110 & 173 to 213 & 358 to 409 & 730 to 802 & \\
& 96 & 192 & 384 & 770 & \\
2.1 & 200,000 & 100,000 & 50,000 & 25,000 & 375,000 \\
& 83 to 109 & 172 to 211 & 358 to 411 & 726 to 807 & \\
& 96 & 192 & 384 & 767 & \\ \hline \hline
\end{tabular}
\end{center}
\end{largetable}

We generate clusters with fixed prefactor ($\kf = 1.3$) and variable fractal dimension
$\df = 1.6, 1.8 , 2.1$. The clusters we analysed belonged to generations
$n = 4$ to $n=7$, their number varying from 200,000 (400,000) to
25,000 (50,000) for $d_f = 1.6, 2.1$ ($d_f = 1.8$). At every generation the
number of clusters decreases by 2. The number of monomers
per cluster varied from approximately 80 to 800, depending
on the generation number. The number of clusters considered,
their size rage, and mode
are shown in table~\ref{table:Agglomerates}.

\section{Appendix C: Discrete fragment size distribution}
\label{sec:AppendixB}

The discrete fragment size distribution is the beta distribution
\beq
p_{ij} \equiv 2 g_{ij} = 2 \, \frac{[i(j - i)]^{\beta(d_f,j)}}
{\sum_{i=1}^{j-1} [i(j - i)]^{\beta(d_f,j)}} ,
\quad i = 1, \ldots j-1 .
\label{eq:DiscreteFit}
\eeq
The exponent $\beta(d_f, j)$ is determined by requiring that
\beq
g_{1j} = \frac{(j - 1)^{\beta(d_f,j)}}
{\sum_{i=1}^{j-1} [i(j - i)]^{\beta(d_f,j)}} 
\label{eq:MonProbDiscrete}
\eeq
equal the probability of monomer occurrence as determined from the empirical
distribution, \emph{i.e.,} from the normalized histogram
of fragment sizes. In fact, as in the case of the continuous distribution,
eq.~(\ref{eq:MonProbDiscrete}) was compared 
to the numerically determined probability of monomer occurrence,
as determined from the fit eq.~(8) in the main text.
\begin{figure}[htb]
\begin{center}
\includegraphics[scale=0.50]{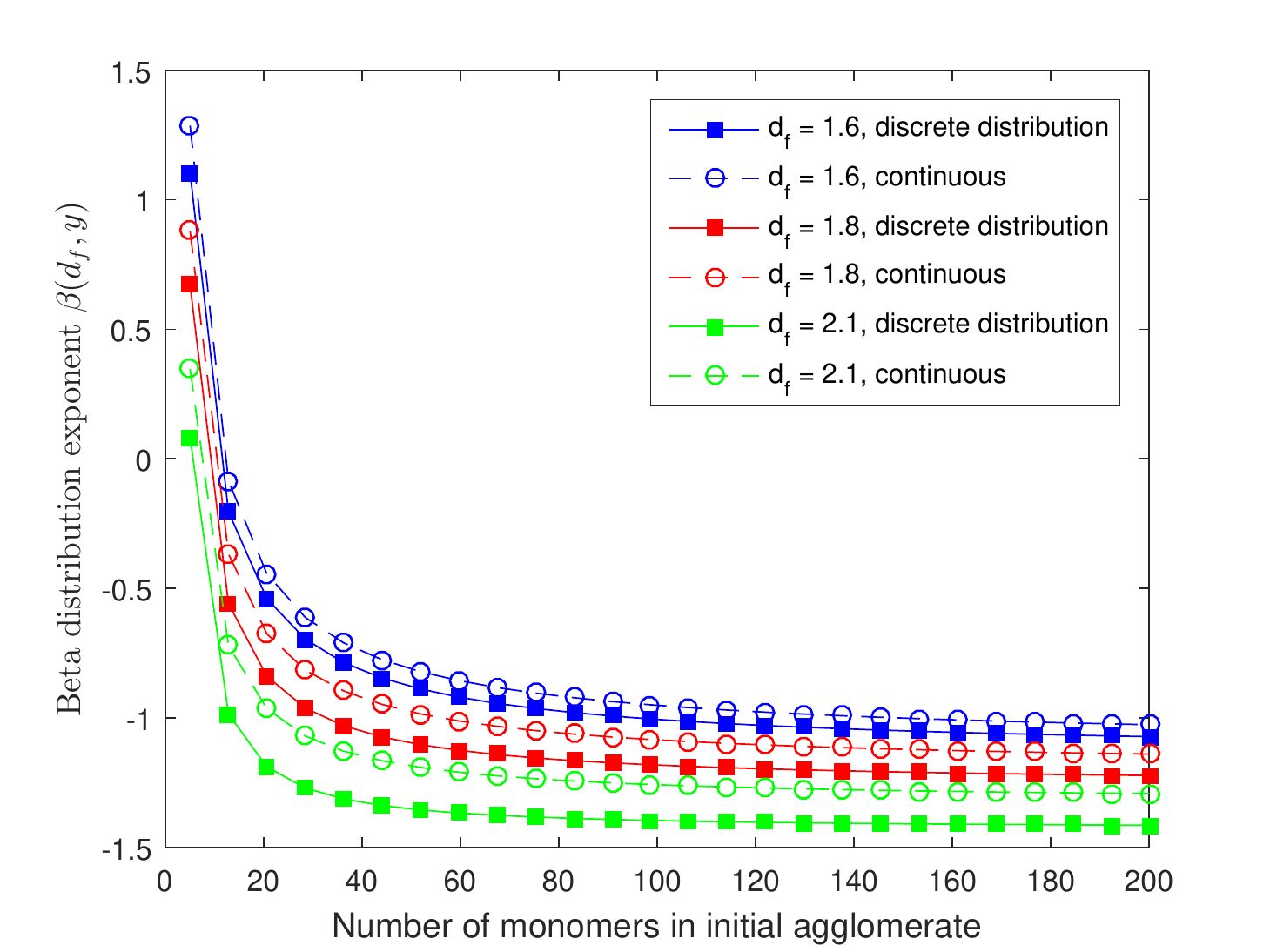}
\caption{Exponent of the discrete $\beta(d_f,j)$ and continuous $\beta(d_f,y)$
beta fragment size distributions as a function of the initial agglomerate size ($j$ or $y$),
parametrized by the fractal dimension (fractal prefactor $k_f = 1.3$). Discrete
distribution (filled symbols), eqs.~(\ref{eq:BetaExpDiscrete}); continuous
distribution (open symbols), eqs.~(10), main text.}
\label{fig:CompareBetaDiscreteCont}
\end{center}
\end{figure}
The highly non linear equation was solved numerically.
The resulting exponent $\beta(d_f,j)$ was fitted to the
non-linear eq.~(\ref{eq:BetaFitSI}), eq.~(10a) in the main text
and reproduced here for completeness,
\begin{subequations} 
\beq
\beta(d_f,y) = a(d_f) + b(d_f) \, j^{c(d_f)}.
\label{eq:BetaFitSI}
\eeq
to obtain the coefficients
\begin{align}
a(d_f) & = -0.56 \, d_f - 0.25; \quad b(d_f) = 4.70 \, d_f + 2.47; \nonumber \\
c(d_f) & = -0.76 \, d_f + 0.29 \quad \mbox{\textrm{for}} \quad
\df \, \epsilon \, [1.6, 2.1].
\label{eq:BetaCoefficientsDiscrete}
\end{align}
\label{eq:BetaExpDiscrete}
\end{subequations}

Figure~\ref{fig:CompareBetaDiscreteCont} compares the calculated exponent of the
beta distribution for the discrete [$\beta(d_f, j)$] and continuous
[$\beta(d_f, y)$] fragment size probability
distributions. The slight differences arise from the normalization condition: a discrete
sum versus a definite integral. 

\end{document}